\documentclass{article}

\usepackage[utf8]{inputenc}
\usepackage{enumitem}
\newcommand{\subscript}[2]{$#1#2$}
\usepackage{changepage}
\usepackage{textcomp,marvosym}
\usepackage{fixltx2e}
\usepackage{amsmath,amssymb}
\usepackage{cite}
\usepackage{nameref,hyperref}
\usepackage[right]{lineno}
\usepackage{microtype}
\DisableLigatures[f]{encoding = *, family = * }
\usepackage{rotating}
\usepackage[aboveskip=1pt,labelfont=bf,labelsep=period,justification=raggedright,singlelinecheck=off]{caption}
\usepackage{tabularx}
\usepackage{booktabs}
\usepackage{authblk}

\title{Understanding Editing Behaviors in\\Multilingual Wikipedia}

\author[1]{Suin Kim\thanks{These authors contributed equally to this work.}}
\author[1]{Sungjoon Park${}^*$}
\author[2]{Scott A. Hale}
\author[1]{Sooyoung Kim}
\author[1]{Jeongmin Byun}
\author[1]{Alice Oh\thanks{alice.oh@kaist.edu}}

\affil[1]{School of Computing, KAIST, Daejeon, Republic of Korea}
\affil[2]{Oxford Internet Institute, University of Oxford, Oxford, UK}

\date{}

\begin{document}

\maketitle

\section*{Abstract}

Multilingualism is common offline, but we have a more limited understanding of the ways multilingualism is displayed online and the roles that multilinguals play in the spread of content between speakers of different languages.
We take a computational approach to studying multilingualism using one of the largest user-generated content platforms, Wikipedia. We study multilingualism by collecting and analyzing a large dataset of the content written by multilingual editors of the English, German, and Spanish editions of Wikipedia. This dataset contains over two million paragraphs edited by over 15,000 multilingual users from July 8 to August 9, 2013. 
We analyze these multilingual editors in terms of their engagement, interests, and language proficiency in their primary and non-primary (secondary) languages and find that the English edition of Wikipedia displays different dynamics from the Spanish and German editions.
Users primarily editing the Spanish and German editions make more complex edits than users who edit these editions as a second language. In contrast, users editing the English edition as a second language make edits that are just as complex as the edits by users who primarily edit the English edition. In this way, English serves a special role bringing together content written by multilinguals from many language editions.
Nonetheless, language remains a formidable hurdle to the spread of content: we find evidence for a complexity barrier whereby editors are less likely to edit complex content in a second language. In addition, we find that multilinguals are less engaged and show lower levels of language proficiency in their second languages.
We also examine the topical interests of multilingual editors and find that there is no significant difference between primary and non-primary editors in each language.

\section{Introduction}

Wikipedia is the world's largest general reference work, and it depends on active editors to generate and maintain up-to-date and accurate information. Wikipedia is also one of the top ten websites in terms of traffic volume, and its articles are often among the top results for many search queries on Google~\cite{lewandowski2011ranking}.

There are 288 language editions of Wikipedia hosted by the Wikimedia Foundation, providing easy access to information for many Internet users globally, but there are high levels of inequality and asymmetry in the information available in the different language editions. 

\subsection{Information Inequality and Asymmetry}

The English edition contains more than 4.9 million articles as of June 2015, which is 13.8\% of all the articles in the 288 language editions~\cite{2015wikistats}. In comparison, the Chinese edition has 827,273 articles and the Arabic edition has 373,064 articles. When looking at the number of edits or active editors, the inequality is even greater. The number of active editors is greatest for the English edition of Wikipedia, which has 6.4 times more active editors than the second most active edition, German. About 38.4\% of all edits ever made to Wikipedia were to the English edition. 
This results in situations where users search for information on Wikipedia, only to find that it is not available in their own languages. For example, searching for PLOS ONE on Wikipedia finds the longest and most comprehensive article in English, but no results in Russian and only very short articles in German and Arabic. In addition to the inequality shown by the large differences in the number of articles and active editors, there is also an asymmetry of information between the different language editions: many topics are available in only one language or a small number of languages~\cite{hecht2010babel}. This applies even to the English edition. Although the English edition is the largest edition, it contains, for example, only 51\% of the articles that exist in the German edition~\cite{hecht2010babel}. Such asymmetry is especially pronounced for Wikipedia articles about local places and events, which are mostly written only in the local languages of those locations~\cite{sen2015}. For example, KAIST is a major science and technology university and research institution in Korea with many international students and faculty, but there is no article about it in the Spanish edition of Wikipedia.

There are several reasons for this information inequality and asymmetry in Wikipedia. First, Wikipedia was only available in English when it started in January 2001. The German and the Catalan editions were added two months later, and other language editions followed after a few years, but English has always remained the largest edition~\cite{2015historywikipedia}. Second, English is a de facto standard language of the Internet and the hub among all global languages~\cite{danet2007multilingual, hale2014wiki, ronen2014links}. Lastly, many editors tend to contribute local content and that leads to an asymmetry in the information available in the different language editions~\cite{hecht2010localness}. For example, on average 23\% of edits by unregistered users of the English edition of Wikipedia were to articles less than 100km from their location~\cite{hecht2010localness}. 

As awareness of this information inequality and asymmetry has increased there have been efforts to grow or distribute more information for several language editions. In 2010, Google sponsored a contest to encourage students in Tanzania and Kenya to contribute to the Swahili edition of Wikipedia~\cite{2010cohenhungry}, and in 2007 the German government allocated funds to support the creation of articles in the German edition of Wikipedia~\cite{2007heisewikipedia}. Even with these efforts, the English edition remains the largest and the most extensive with 2.5 times more articles compared to the second-largest edition, Swedish.

\subsection{Multilingual Editors in Wikipedia}

Previous research has shown that multilingual users play a key role in information diffusion across languages in social media~\cite{herring2007,kim2014sociolinguistic,eleta2012,hale2014twitter}. 
In Wikipedia, where editors are not explicitly linked in a social network, multilingual editors can still play a key role in mitigating the asymmetry by transferring information across languages~\cite{hale2014wiki}. Thus, we present in-depth research to understand the editing behaviors of multilingual editors in Wikipedia.

Approximately 15\% of active Wikipedia editors are multilingual, contributing content to multiple language editions of the encyclopedia~\cite{hale2014wiki}. Wikipedia provides a global account system for a single login across all Wikimedia sites, as well as interlanguage links connecting articles on the same concepts across languages.
However, other than a small mixed-methods study of the contributions of Japanese--English bilingual editors on articles about Okinawa, Japan~\cite{hale2015}, little is known about the content contributions of these multilingual editors at a larger scale or across different language pairs.

Unlike the studies on monolingual Wikipedia editors~\cite{yasseri2012practical, ortega2008inequality, iba2010analyzing, lieberman2009you}, we must consider that a multilingual editor's behavior may differ for each language. Studies have shown that multilingual speakers feel differently in each of the languages they speak~\cite{Pavlenko2005, Dewaele2010} and tend to use a different language depending on the purpose, domain, and conversational partner~\cite{Barron2011}. In addition, multilingual individuals rarely possess equal and perfect fluency in all their languages~\cite{haugen1969}. Given this, we expect multilingual Wikipedia editors to also behave differently in each language edition of Wikipedia they edit. More specifically, they may contribute to one language more than another, they may edit articles on different topics in each language, and their edits may demonstrate different levels of proficiency in each language. While comparing multilingual editors' behaviors within each language may be ideal, different languages cannot be compared directly in terms of topics and proficiency because of inherent differences among them.
Instead, we look at the behaviors of the multilingual editors within each language. For each language edition, we group the editors into those who edit that language edition more than any other language edition (\textit{primary  editors} of that language) and those who edit some other language edition more (\textit{non-primary editors} of that language) and then compare the behaviors of the primary and non-primary multilingual editors. We then ask the following three research questions about multilingual editors in the English, German, and Spanish language editions.
\begin{enumerate}[label=\subscript{RQ}{\arabic*}.]
\item Do primary and non-primary editors show different levels of \textit{engagement}?
\item Do primary and non-primary editors show different levels of topical \textit{interest}?
\item Do primary and non-primary editors show different levels of \textit{language proficiency}?
\end{enumerate}
Each editor's engagement is quantitatively measured by their text contributions and the time spent revising articles. The interests of editors are identified by the topics of the articles they edited. Lastly, the language proficiencies of editors are measured by various language complexity measures applied to their contributions.

Our results can be summarized in three parts. 
First, there is are significant differences in the levels of engagement and language proficiency in the German and Spanish editions: multilingual users who primarily edit either of these editions show higher levels of engagement and language proficiency than multilingual users who edit these editions as non-primary languages. Second, there was no notable difference between the degrees of engagement and proficiency of editors who edited the English edition as a primary or non-primary language, verifying the common assumption of English as lingua franca of the Web. 
Third, primary and non-primary editors show similar levels of interest for most topics with the exception that primary editors are significantly more interested in local topics and non-primary editors in global topics.

Our contributions to the field of Web-based study of multilingualism are as follows:
\begin{itemize}
    \item We construct an extensive dataset of multilingual Wikipedia edit history, which comprises more than 5 million edits and make it publicly available for future research.
    \item We define and analyze three relevant aspects of multilingual editors' behavior: engagement, interest, language complexity.
    \item We define and validate several language-independent measures for quantifying the language complexity of edits.
    \item We show that multilingual editors indeed have potential to help mitigate the inequality and asymmetry in the information available in different languages.
\end{itemize}

\section{Methods}

In this section, we describe the data collection and analysis methods. To analyze the editors' engagement, interests, and language proficiency levels, we first start with the edit metadata from the English, German, and Spanish editions for one month and construct article edit sessions to identify consecutive contributions to articles by the same editor. Based on these article edit sessions, we extract multilingual editors and their contributions. By analyzing their contributions, we (1) measure their degrees of \textit{engagement} through the number of contributions they made to articles, (2) discover their \textit{interests} through the topics of the articles they edited, (3) estimate their \textit{language proficiency} levels for each language by measuring the language complexity of their edits.

\vspace{2mm}

\subsection{Dataset}
To extract only the contributions of multilingual editors to articles from the unstructured edit history data, we conduct a data processing pipeline consisting of the following steps. We first start with metadata of English, German, and Spanish language editions from July 8 to August 9, 2013. We then construct article edit sessions by grouping together consecutive edits to the same article by the same editor. Based on the identified sessions, we define multilingual editors as those who are involved in article edit sessions for two or more language editions. Finally, we download the actual revision text for the article edit sessions of those multilingual editors. We describe these steps in more detail below.

\paragraph{Identifying Edits from Metadata.}
We begin with the edit metadata collected by Hale~\cite{hale2014wiki} for the largest 46 language editions of Wikipedia and extract the data for English (en), German (de), and Spanish (es). The metadata includes article titles, language editions, timestamps, editor ids, and URLs to the content of each edit captured from a near real-time broadcast on Internet Relay Chat. The extracted data comprises 2,799,729 edits by 146,616 distinct editors. We discard edits to non-article pages including article talk pages and user pages. We further discard edits that are indicated as being made by Wikipedia (ro)bots. In contrast to Hale~\cite{hale2014wiki}, we retain edits marked as ``minor'' because what one considers as minor may vary from person to person and language to language.

\paragraph{Article Edit Session.} Even though the easiest way to measure the edit activity on Wikipedia is simply counting each edit when changes are submitted, this does not accurately reflect editors' behavior because of individual differences in activity patterns. For example, some editors may submit a few large edits while others may make a series of smaller edits saving the pages more frequently as they work. To account for these individual differences, we adopt the idea of \textit{edit sessions}~\cite{geiger2013using}, which measure the labor of Wikipedia editors whose contributions tend to occur in bursts. In our work, we limit each edit session to a single document, and so we rename it \textit{article edit session}. We use \textit{one hour} as the cutoff between intra-session and inter-session edit activities. 

In other words, from the collected dataset, we define an \textit{article edit session} as a sequence of continuous editing activity without a break of more than one hour for a specific article by a single editor. 

\vspace{2mm}

\paragraph{Primary vs. Non-Primary Editors.}
Next, we identify multilingual editors from the metadata and retrieve the content of all the edits made by multilingual editors from Wikipedia using the Wikipedia API. We define a multilingual editor as an editor who edits two or more language editions. Using this definition, we identified 12,577 multilingual editors with 427,529 total article edit sessions composed of 622,766 distinct edits. Following Hale~\cite{hale2014wiki}, we operationally define primary and non-primary users as follows:
\begin{itemize}
\item 
We identify the \textit{primary} language of an editor as the language of the edition that the multilingual editor edits most frequently.
\item
We identify the \textit{non-primary} languages of an editor as all the other languages that they edit.
\item
For each language edition X, we define \textit{primary editors} as the set of all users with primary language X.
\end{itemize}
Thus, an editor can be a primary editor in only one language edition, but can be a non-primary editor in multiple language editions.

\begin{figure}
\includegraphics[width=1\textwidth]{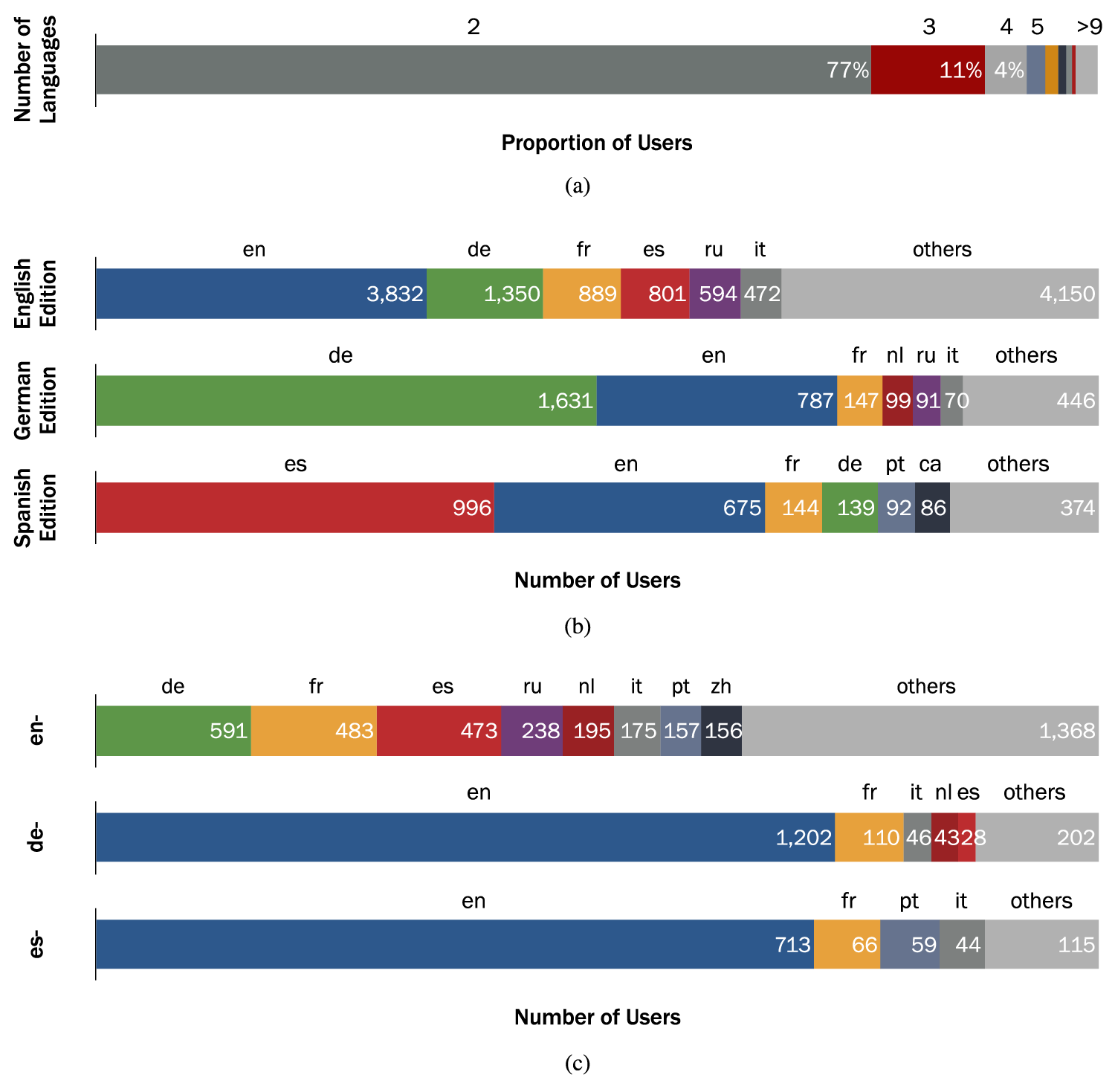}
\caption{(a) {\bf Number of languages multilinguals edit in Wikipedia.} Regarding the number of languages they edit, 77.3\% of multilingual editors are bilingual, followed by 11.4\% trilingual and 4.1\% and quadrilingual editors. We discard editors who edited more than 10 languages (these editors account for 2.3\% of all multilingual editors). (b) {\bf Primary languages of multilingual editors for the three largest language editions.} The English edition has the largest, yet the most varied, number of multilingual editors by primary language. 32.9\% of the multilingual editors who edited the English edition primarily edited English. In comparison, 49.9\% of the multilingual editors in the German edition primarily edited the German edition. Multilingual editors who primarily edit English are the second-largest proportion of multilingual editors in the Spanish and German editions. (c) {\bf Second most used languages for the three primary editor groups.} English primary editors have much more diverse language usage, compared to German and Spanish primary editors. Most of German and Spanish primary editors contribute to English edition as their second most edited version. }
\label{numlang_userproportion}
\end{figure}

Fig.~\ref{numlang_userproportion}a shows the distribution of the number of languages per user in our multilingual editor dataset. We found that most multilingual editors contribute to two or three language editions. There are also a small number of editors with edits in more than 10 languages, including two who edited all 46 languages we studied. We regard all editors with more than 10 editions as either outliers or Wikipedia bots not setting the `bot' flag for their edits and drop them from further study.
Table~\ref{data_statistics} contains the statistics of the edit history data we use, and Fig.~\ref{numlang_userproportion}b shows the proportions of the primary languages of editors who are contributing to the English (en), German (de), and Spanish (es) editions.

\paragraph{Collecting Edit Text.}
Wikipedia provides the previous and current versions of each article. To look at the actual edited text, we download the diffs from the Wikipedia Web interface for each article edit session we identified. For each article edit session, we extract the pairs of changed paragraphs and convert the edit from Wiki markup to plain text. 
In this way, we retain only the visible text from edits, discarding all non-visible and non-text information including URL, multimedia, metadata, and document structure. We regard an edit as \textit{non-visible} if there is no visible text change. An example of such an edit is adding a link to existing text but otherwise making no other changes.

\begin{table}[tb]
\begin{center}
\begin{tabular}{ lrrrrrr } 
\toprule
& en-p & en-np & de-p & de-np & es-p & es-np \\
\midrule
\# Multilingual Editors & 3,832 & 7,784 & 1,631 & 1,640 & 996 & 1,510\\
\# Article Edit Sessions & 200,883 & 36,959 & 112,788 & 7,334 & 63,947 & 5,609\\ 
\# Edits & 298,868 & 51,665 & 151,014 & 9,111 & 104,341 & 7,757\\
\# Edited paragraphs & 1,447,692 & 230,893 & 816,647 & 27,656 & 554,762 & 25,340\\
\bottomrule
\end{tabular}
\vspace{5mm}
\caption{ {\bf Number of editors, article edit sessions, edits, and edit paragraphs.} 
There is more activity in the English edition (en) than in either the German (de) or Spanish (es) edition. In all three language editions there are more primary editors (p) than non-primary editors (np) and primary editors are more active than non-primary editors.
}
\label{data_statistics}
\end{center}
\vspace{5mm}
\end{table}

\begin{figure}[!htb]
\centering
\includegraphics[width=1\textwidth]{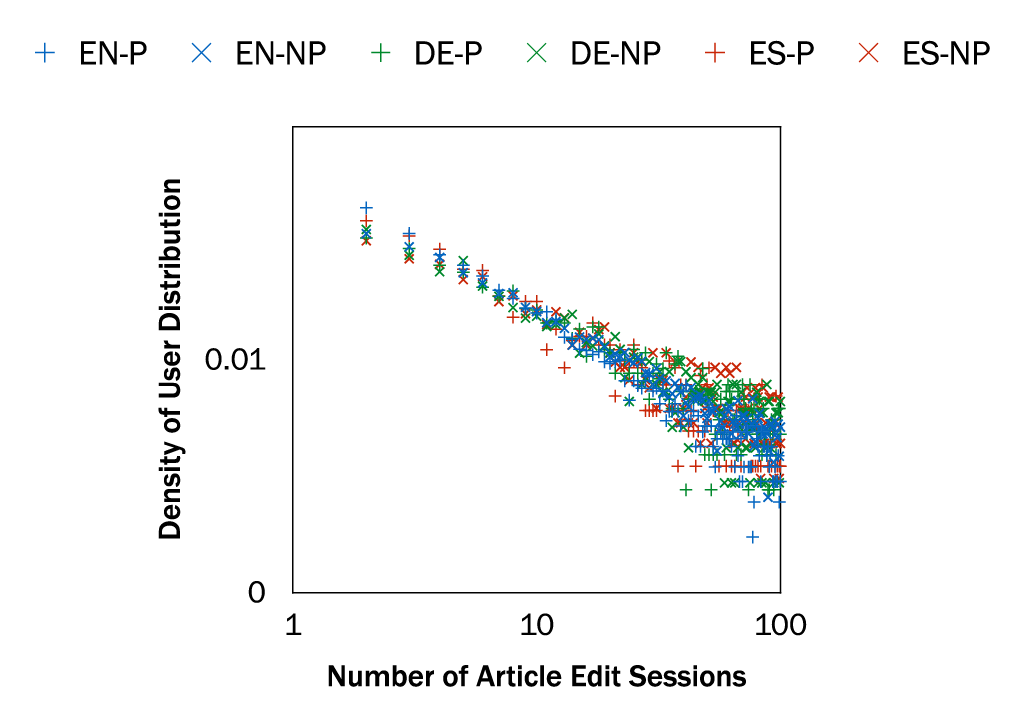}
\caption{{\bf Distribution of editors and article edit sessions for the three language editions---English (en), German (de), and Spanish (es)---for primary (P) and non-primary (NP) multilingual editors.} We display the number of article edit sessions up to 100. Dots in the same color denote the same language edition. The plot, on a log-log scale, shows that the distributions for primary and non-primary users in all three editions is heavy-tailed; most users perform only a few edits while a few users perform many edits.}
\label{languagegroup_numusers}
\end{figure}

\paragraph{Distribution of Article Edit Sessions}
We examine the distribution of multilingual editors' contributions by language and the number of article edit sessions. Fig.~\ref{languagegroup_numusers} shows the distribution of editors by the number of article edit sessions. The plot, on a log-log scale, shows that the distributions for primary and non-primary users in all three editions is heavy-tailed; most users perform only a few edits while a few users perform many edits. 

\subsection{Quantifying Editors' Behavior} 
Our approach is to analyze and compare the behavior of primary and non-primary multilingual editors for the English, German, and Spanish editions in terms of \textit{engagement}, \textit{interest}, and \textit{language proficiency}. We measure editors' engagement by looking at the editing patterns quantitatively within article edit sessions. Then, we investigate editors' interests by focusing on the content of edits in terms of the main topics of the articles they edited. Lastly, we measure levels of language proficiency by defining and computing several language complexity metrics.

\begin{figure}[!htb]
\centering
\includegraphics[width=1\textwidth]{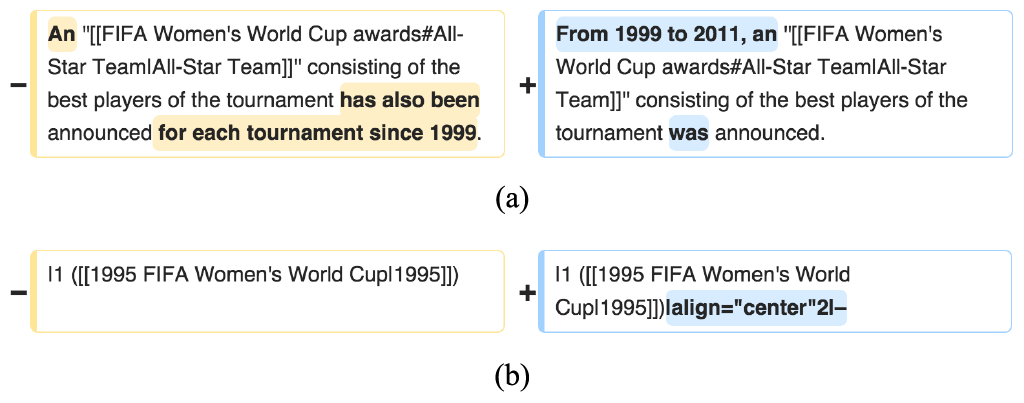}
\caption{{\bf Example of an edit paragraph in a Wikipedia article 2015 FIFA Women's World Cup.} (a) Edit paragraph containing a visible edit. (b) Edit paragraph containing a non-visible edit. We define an edit paragraph as one line of Wikipedia markup, utilizing the diff from Wikipedia. \texttt{Pre-edit} text and \texttt{post-edit} text are shown on the left and right sides, respectively, with text differences highlighted.}
\label{wikipedia_paragraph_example}
\end{figure}

\paragraph{Examining Edits in Four Aspects.}
We define an edit paragraph as a line of Wiki markup in an article, utilizing the difference of the text before and after the edit from Wikipedia as shown in Fig.~\ref{wikipedia_paragraph_example}. In an article edit session, all of the edit paragraphs before revisions, shown in the left side of Fig.~\ref{wikipedia_paragraph_example}, are paired with their corresponding revised edits, shown in the right side of Fig.~\ref{wikipedia_paragraph_example}. With those paired paragraphs, we examine the edit paragraphs in four ways as follows:

\begin{itemize}
    \item {\bf Pre-edit}: Edit paragraph (text) before revision
    \item {\bf Post-edit}: Edit paragraph (text) after revision
    \item {\bf Diff}: The actual revision including both inserted and deleted text in an edit
    \item {\bf Delta}: The difference between a \texttt{Pre-Edit} paragraph and its corresponding \texttt{Post-Edit} paragraph for each computed measure
\end{itemize}

The \texttt{Diff} includes both inserted and deleted text and is used to measure an editor's direct contributions. The difference between computed measures for \texttt{Pre-edits} and \texttt{Post-edit}s, defined as \texttt{Delta}, is an indirect approach that considers the context of an edit.

\paragraph{Comparing Editor Behaviors within Languages}
Rather than comparing each editor's behavior across languages, we compare editing behavior of primary versus non-primary editors within each language, as defined in the previous section. That means, for example, we take all editors in our dataset who contributed to any articles in the English edition, divide them into two groups: 1) those who contributed to the English edition more than any other language edition, and 2) those who contributed to some other language edition more than the English edition, and then we compare the behaviors of those two groups. This way, we do not need to worry about the inherent differences among languages (e.g., German tends to have longer words than English). 

\paragraph{Editor Engagement.}
Since we categorized multilingual editors into two groups (primary and non-primary) on the basis of the number of article edit sessions they had across language editions, we measure the level of quantitative engagement of editors within an article edit session, such that the measure becomes independent of the number of article edit sessions. 

We measure the amount of engagement based on four metrics. First, we count the number of revisions committed within an article edit session, only including the sessions containing more than 1 edit. Second, we measure the session length in minutes by the difference of timestamps between the first and the last revision in each article edit session. Third, we measure the amount of text added and deleted in terms of characters, words, and sentences for all the \texttt{Delta} in a session. Last, we count the number of edit sessions that only include non-visible edits indicating the changes the editor made did not affect the text of the article (i.e, modifying URL link, images, etc.). By computing this, we can examine whether an editor's engagement results in article content changes. These four metrics are summarized in Table~\ref{tab:measures}.

For each metric, we compute the mean over all article edit sessions for each editor, and then again compute the mean of primary and non-primary editors to represent the level of engagement for each groups. We perform two-tailed independent samples t-test to examine if the differences between the means of the two groups are significant.

\paragraph{Editor Interest.}
Assuming that the topics of the edited articles represent editors' interests in specific fields, we measure the interests of multilingual editors using a Bayesian topic model. We first determine the main topic category for each article a multilingual user edited and assign each article a single topic label. We then compute the proportion of interest from primary and non-primary editors for each topic. By comparing the distributions, we can compare the interests of primary and non-primary editors.
\begin{itemize}
    
    \item \textbf{Topic Modeling.} We use a Bayesian topic model as an automatic method to categorize Wikipedia articles into different topics. Although Wikipedia already provides ``categories", there are many cases where one Wikipedia article is assigned to multiple categories. Moreover, Wikipedia categories form a complex network in which the relationships between categories are often unclear~\cite{kittur2009}. Instead of using these Wikipedia-defined categories, we develop a more consistent and replicable methodology. We first model the topics using a widely-used topic model, the latent Dirichlet allocation (LDA)~\cite{Blei03latentdirichlet} using the Python Gensim~\cite{rehurek_lrec} library with the online variational Bayes algorithm. We set the number of topics ($k$) to 100 and use the default values for the hyperparameters.

    \item \textbf{Clustering and Labeling Articles.} With the 100-dimensional topic proportion vectors generated by LDA from the step above, we cluster the articles using DBSCAN~\cite{ester1996density} in the Python Scikit-learn~\cite{scikit-learn} package. The generated clusters show consistency in the topics of the articles; hence, we consider each cluster as representing a single topic. We manually examine the articles in each cluster, assign a topic label to the cluster, and assign the same topic label to each article in the cluster. This process results in 20 clusters and their topic labels for each language edition with all articles labeled with one of the 20 topic labels. To verify the clustering process, we simply compare the inter-cluster and intra-cluster distances using Euclidean distance. In Eq.~\ref{eq:Ici}, the numerator is the average of all pairwise distances of the articles in a cluster, representing the intra-class distance. The denominator is the mean of the pairwise distances of the cluster medoids for all clusters, representing the inter-class distance. For both the numerator and the denominator, $c_i$ denotes cluster $i$ in the set of all clusters $C$, and $x$ denotes each article in $c_i$. The resulting $I_{c_i}$ represents the ratio of intra-class to inter-class distance; so, we would expect this ratio to be significantly below one if the clustering is done well. The average value of $I_{c_i}$ for $i$th cluster is 0.59 with a standard deviation 0.22. 

\begin{equation}
    \mathrm{medoid}(c_i)=\arg\!\min_m\sum_{x\in{c_i}}{dist(x, m)}
    \label{eq:medoid}
\end{equation}
\begin{equation}
    I_{c_i}=\frac{\frac{1}{\left\vert{c_i}\right\vert}\sum_{x\in{c_i}}{\mathrm{dist}(x, \mathrm{medoid}(c_i))}}{\frac{1}{\left\vert{\mathrm{C}}\right\vert - 1}\sum_{{c_j}\in{\mathrm{C}}, c_j \neq c_i}{\mathrm{dist}(\mathrm{medoid}(c_i), \mathrm{medoid}(c_j))}}
    \label{eq:Ici}
\end{equation}

 \item \textbf{Comparing Levels of Interest.} The previous step provides topic labels for each article. With those labels, we can define, for each multilingual editor, the level of interest on a topic as the proportion of the article edit sessions for articles in that topic over all article edit sessions by the editor. Then, we average the interest levels of editors in the primary and non-primary groups. We define this as the ``normalized frequency of edits for each topic" (also listed in Table~\ref{tab:measures}). To compare the interest levels between the two groups, we perform a two-tailed independent samples t-test for each topic.
 
 \end{itemize}

\paragraph{Language Proficiency.}
In order to measure the language proficiency of multilingual editors, we focus on three aspects of their edits: (1) \texttt{Pre-Edits}, (2) \texttt{Delta}, and (3) \texttt{Diff}.

Using \texttt{Pre-Edits}, we analyze the language complexity of the paragraphs of the articles that multilingual editors choose to revise. Using \texttt{Delta} and \texttt{Diff} we quantify the editors' contributions in terms proficiency to estimate editors' linguistic abilities. 
With \texttt{Pre-Edits} and \texttt{Delta}, the textual contents are sufficient in length to perform both lexical diversity and syntactic complexity measures, which have previously been used as estimates of language proficiency in language acquisition research~\cite{bulte2014conceptualizing}. Meanwhile, for \texttt{Diff}, which often consist of just a few words, we focus on the definite and indefinite articles edited (e.g., `a', `an', and `the' in English). These have also been shown by previous research to be a good proxy for language proficiency in English, German, and Spanish~\cite{butler2002second, ibanez2014annotating, jaensch2008l3}.

\begin{itemize}

\item \textbf{Lexical Diversity Measures.}
We first compute the entropy of unigram, bigram, and trigram frequencies as complexity measures~\cite{Manning:1999:FSN:311445}. These measures are used to quantify the richness of word usage in Wikipedia articles~\cite{yasseri2012practical}. To calculate the entropy for each edit paragraph, we average the metrics over all unigrams, bigrams, and trigrams that appear in the paragraph.

\item \textbf{Syntactic Complexity Measures.}
We compute the entropy of parts-of-speech (POS) frequency. Analyzing the sequence of POS has advantages over analyzing the sequence of words. POS entropy can ignore extremely trivial edits (e.g., correcting typos) as well as meaningless bot-produced edits~\cite{yasseri2012practical}. In addition, looking for diverse combinations of POS is a good approach for detecting complex syntactic structure~\cite{BrianRoark2007}. We count the occurrence of POS unigrams, bigrams, and trigrams in each edit paragraph and then compute an entropy measure based on the frequency distributions of POS that captures the amount of information and thus indicates the diverse use of POS in the edits. 

For automatic POS tagging on edits, we employ a maximum entropy POS tagger~\cite{ratnaparkhi1996maximum} trained on the Penn Treebank corpus for English edits with Penn Treebank tagset~\cite{Marcus93buildinga}. For German edits, we use the Stanford log-linear POS tagger~\cite{Toutanova:2003:FPT:1073445.1073478} trained on the NEGRA corpus with Stuttgart-T{\"u}binger Tagset (STTS)~\cite{Skut98alinguistically}. For Spanish edits, we use the tagger trained on the AnCora corpus with its tagset~\cite{Taule08ancora:multi}. 

\item \textbf{Usage of Articles.} A number of previous studies investigate the difficulty of learning the definite and indefinite articles for language learners. For example, Butler~\cite{butler2002second} finds that children who are learning English as their second language (L2) have more difficulty in understanding articles compared to children who are acquiring English as their first language (L1). They show that L1 learners make lower frequency of errors than L2 learners. In addition, Japanese-speaking English learners have the most difficulty in understanding the article system and
even proficient English learners only record about a 90\% of success rate when given a task to choose the proper article~\cite{mcenery2006corpus}. In addition, the difficulty of the article system in Spanish and German are comparable to English~\cite{ibanez2014annotating,jaensch2008l3}. 

The difficulty comes from the complex usage of articles: i.e., that there is not a one-to-one correspondence between languages. Such difficulty imposes a challenge for language learners of a second language \cite{butler2002second}. Since it is computationally challenging to automatically test the appropriateness of the articles editors use, we approximate the level of understanding of the article system in each language by measuring the frequencies with which definite and indefinite articles are used. We count the number of definite and indefinite articles in each \texttt{Diff} by primary and non-primary users. The definite and indefinite articles we count in each language are shown in Table~\ref{language_article_list}. The number of articles in a \texttt{Diff} is divided by total number of words of the \texttt{Diff}, and we refer to this quantity as the ``fraction of articles in added tokens.''

\end{itemize}

\begin{table}[!htb]
\centering
    \begin{tabular}{llll}
    \toprule
               & English & German                                & Spanish          \\
    \midrule
    Definite   & the     & des, die, den, der, dem, das          & el, la, los, las \\
    Indefinite & a, an   & eine, eines, einer, einem, einen, ein & un, una, unos, unas \\
    \bottomrule
    \end{tabular}
\caption{\bf The definite and indefinite articles tracked in English, German, and Spanish.}
\label{language_article_list}
\end{table}

\begin{itemize}
\item \textbf{Computing Language Complexity Measures for Each Editor.}
The language proficiency measures were summarized with a \textit{maximum} value for each editor rather than the mean value. We compute the maximum value of the proficiency measures for all edits belonging to the same editor as a representation of that editor's highest displayed level of linguistic ability to produce complex edits. The maximum value is a fairer estimate than the mean or the minimum value as not all possible revisions to an article always require an editor's maximum linguistic ability. For this reason, widely used central tendency measures (e.g., mean) would not reflect editors' true proficiencies properly. Summarizing each editor's edits with the maximum value assumes that each editor shows his/her maximum linguistic ability to edit a paragraph at least once, which is more reasonable. 

Since the maximum value may be merely affected by the number of edits, we control the number across editors to three. We uniformly sample three edits within the entire edits an editor made, repeating 100 times for editors and average the evaluated 100 maximum values.

\end{itemize}

All of the engagement, interest, and proficiency measures described in this section are summarized in Table~\ref{tab:measures}.

\begin{table}
\begin{tabular}{p{36.5em}}
	\toprule\vspace{-5mm}
	\begin{itemize}
      \setlength\itemsep{-0.5em}\setlength\itemindent{-2em}

      \item \textbf{Engagement}
      	
      	\begin{itemize}\vspace{-2.5mm}
        \setlength\itemsep{-0.3em}\setlength\itemindent{-2em}
            \item Number of edits
            \item Session length in minutes
            \item Number of edited characters, words, sentences (for \texttt{Delta})
            \item Fraction of non-visible article edit sessions.
      	\end{itemize}

      \item \textbf{Interest}

      	\begin{itemize}\vspace{-2.5mm}
        \setlength\itemsep{-0.3em}\setlength\itemindent{-2em}
            \item Normalized frequency of edits for each topic
        \end{itemize}

      \item \textbf{Language Proficiency}        
            
      	\begin{itemize}\vspace{-2.5mm}
        \setlength\itemsep{-0.3em}\setlength\itemindent{-2em}
            
            \item Lexical diversity measures (for \texttt{Pre-Edit} and \texttt{Delta})
          	\begin{itemize}\vspace{-1.5mm}
          	\setlength\itemsep{-0.1em}\setlength\itemindent{-2em}
                \item Entropy of unigram, bigram, trigram frequencies
            \end{itemize}
            
            \item Syntactic complexity measures (for \texttt{Pre-Edit} and \texttt{Delta})
          	\begin{itemize}\vspace{-1.5mm}
          	\setlength\itemsep{-0.1em}\setlength\itemindent{-2em}
                \item Entropy of POS unigram, bigram, trigram frequencies
            \end{itemize}
            
            \item Article usage measures (for \texttt{Diff})
            \begin{itemize}\vspace{-1.5mm}
          	\setlength\itemsep{-0.1em}\setlength\itemindent{-2em}
                \item Fraction of articles in added tokens
            \end{itemize}
            
        \end{itemize}

      \vspace{-5mm}
    \end{itemize} \\
    \bottomrule
\end{tabular}
\caption{\bf Measures of engagement, interest, and language proficiency.}
\label{tab:measures}
\end{table}

\section{Results}
In this section, we report our findings on engagement, interests, and language proficiency of primary and non-primary multilingual editors for the English, German, and Spanish editions of Wikipedia.

\subsection{Editor Engagement}

\begin{figure}[!htb]
\centering
\includegraphics[width=1\textwidth]{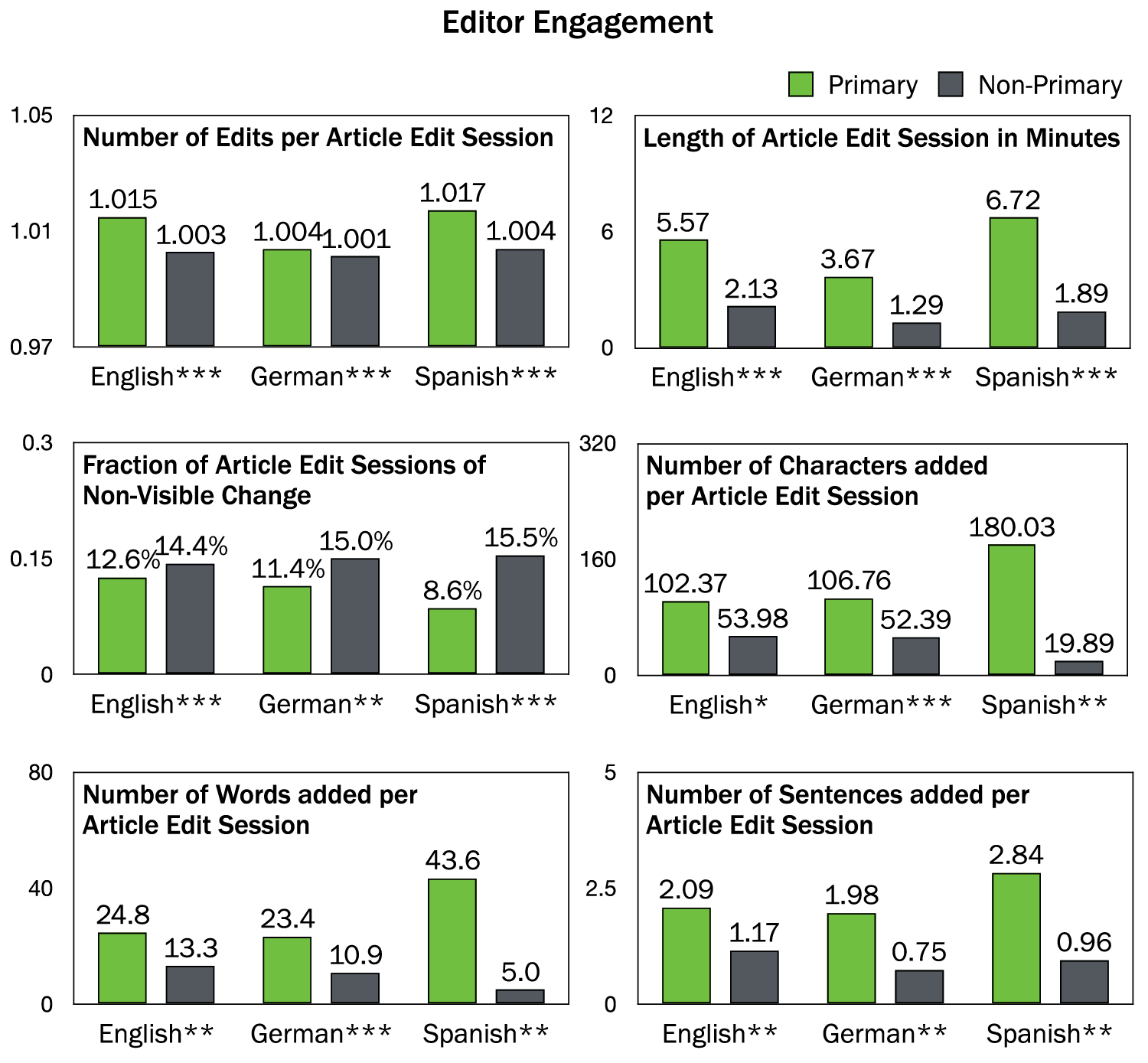}
\caption{{\bf The evaluated engagement measures for primary and non-primary editor groups.} For all metrics and languages, primary and non-primary editors are showing significantly different behavior: primary editors tend to be more engaged than non-primary editors. 
(*$p < 0.05$; **$p < 0.01$; ***$p < 0.001$).}
\label{editstats_documents}
\end{figure}

We show in Fig.~\ref{editstats_documents} that primary editors commit more edits than non-primary editors within an article edit session, for all three language editions. Likewise, the article edit sessions of primary editors are longer than those of non-primary editors. These results indicate that primary editors are more engaged. They  make more edits and spend more time revising each article. 

We also find that the number of tokens added per edit session by primary editors is higher than the number added by non-primary editors in all three language editions. This also holds whether tokens are measured by characters, words, or sentences. These findings on the amount of edited content align with the previous findings on the number of edits and the overall time spent indicating that in general primary editors are more engaged in revising the text of articles than non-primary editors.

Finally, we find that non-primary editors make more non-visible edits, such as adding/removing hyperlinks or applying stylistic changes, in all three languages. This tendency indicates that editors may be making different types of edits in their primary and non-primary languages. Similar results have been shown in qualitative research (e.g., \cite{hale2015}) at a much smaller scale.

\subsection{Editor Interests}
\begin{figure}[!htb]
\centering
\includegraphics[width=0.9\textwidth]{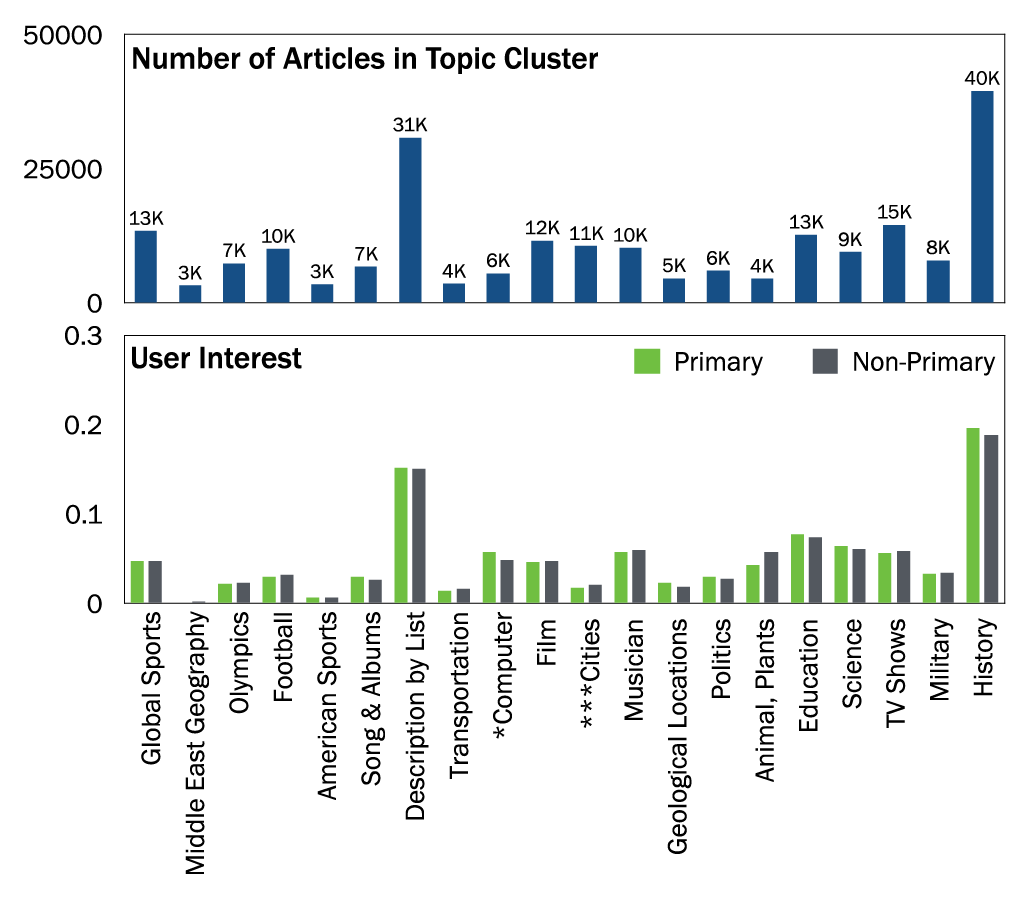}
\caption{{\bf Top: Number of English Wikipedia articles in each topic cluster.} Articles related to Descriptive and History make up a large proportion of all articles while Middle East Geography, American Sports, and Transportation articles are relatively small in proportion. {\bf Bottom:  Differences in interest in the English edition.} Primary and non-primary editors show similar levels of interest for most topics. (*$p < 0.05$; **$p < 0.01$; ***$p < 0.001$)}
\label{topicclusters_all}
\end{figure}

The twenty topics discovered for the articles edited by multilingual users in the English edition are as follows: \texttt{Science}, \texttt{Football}, \texttt{Film}, \texttt{Middle East Geography}, \texttt{American Sports}, \texttt{Songs \& Albums}, \texttt{Musicians}, \texttt{Cities}, \texttt{Global Sports}, \texttt{TV Shows}, \texttt{Politics}, \texttt{History}, \texttt{Military}, \texttt{Transportation}, \texttt{Computer}, \texttt{Education}, \texttt{Geographical Locations}, \texttt{Descriptive}, \texttt{Olympics}, and \texttt{Animals \& Plants}. Fig.~\ref{topicclusters_all} shows the number of articles in each topic in the English edition. The titles of the representative articles for each topic are given in the supporting information~\nameref{appendix:topic_clusters}. 

Similarly, the twenty topics discovered for Spanish articles are as follows: \texttt{Art}, \texttt{Descriptive}, \texttt{Soccer}, \texttt{Film}, \texttt{Animal}, \texttt{Global Sports}, \texttt{History}, \texttt{Plants}, \texttt{Politicians}, \texttt{Natural Science}, \texttt{Social Science}, \texttt{Music}, \texttt{Cities}, \texttt{Geographical Locations}, \texttt{Olympics}, \texttt{Literature}, \texttt{Musicians}, \texttt{Politics}, \texttt{Entertainment}, and \texttt{Tennis}. 

Finally, the twenty topics discovered for German articles are as follows: \texttt{Computer}, \texttt{Natural Science}, \texttt{Descriptive}, \texttt{Geographical Locations:\,U.S.}, \texttt{Names}, \texttt{Geographical Locations:\,Europe}, \texttt{History}, \texttt{Academic}, \texttt{Celebrities}, \texttt{Soccer}, \texttt{Cultural Heritage}, \texttt{Musicians}, \texttt{Natural Topography}, \texttt{Land Transport}, \texttt{Politicians}, \texttt{Entertainment}, \texttt{Air transport}, \texttt{Global Sports}, \texttt{Authors}, and \texttt{Military}.

The overall difference of interest between primary and non-primary editors is not significant ($\chi^{2}_{en}(19)$=0.005, ~$\chi^{2}_{es}(19)$=0.031, ~$\chi^{2}_{de}(19)$=0.019, \textit{ns}). Indeed, as shown in Fig.~\ref{topicclusters_all} bottom, we observe that primary and non-primary editors of the English edition have similar interests. This tendency is also shown in the other language editions.

However, when looking at each topic, we observe notable differences of the level of interests between primary and non-primary editors for a few topics in each language. In the English edition, we observe two topics with significantly different levels of interest between primary and non-primary editors. Primary editors show significantly higher level of interest in the topic of \texttt{Computer}, while non-primary editors show higher level of interest for \texttt{Cities}. In the German edition, primary editors show more interest in \texttt{Computer} and \texttt{Natural Science}, while non-primary editors show more interest in \texttt{Soccer} and \texttt{Global Sports}. In the Spanish editions, primary editors show more interest in \texttt{Politicians}, \texttt{Social Science}, and \texttt{Entertainment}, while non-primary editors show more interest in \texttt{Plants} and \texttt{Geographical Locations}.

In the German and the Spanish editions, we also observe an interesting pattern where topics with more interest from primary users have higher syntactic complexity compared to topics with more interest from non-primary users (see Table \ref{topic_interest_complexity}). For example, in the German edition, \texttt{Natural Science} has an average entropy value of 3.52 compared to 2.52 of \texttt{Soccer}. We do not observe such a pattern in the English edition.

\begin{table}[!htb]
\centering
    \begin{tabular}{llll}
    \toprule
                & English & German                                & Spanish          \\
    \midrule
    Primary     & \texttt{Computer} & \texttt{Computer} (3.21), & \texttt{Politicians} (3.19), \\
                & (2.58)           & \texttt{Natural} & \texttt{Social Science} (3.01),\\
                & & \texttt{Science} (3.52) & \texttt{Entertainment} (2.97)\\
    \midrule
    Non-Primary & \texttt{Cities}   & \texttt{Soccer} (2.76), & \texttt{Descriptive} (2.80),   \\ 
                & (2.63)           & \texttt{Global} & \texttt{Plants} (2.37), \\
                & & \texttt{Sports} (2.52) & \texttt{Geographical} \\
                & & & \texttt{Locations} (2.53) \\
    \bottomrule
    \end{tabular}
\caption{{\bf Topics having significant difference of interest levels between primary and non-primary users in each language edition.} Number next to each topic represents the average entropy for Part-of-Speech trigrams for the articles within each topic.}
\label{topic_interest_complexity}
\end{table}

\subsection{Language Proficiency}
\begin{figure}[!htb]
\centering
\includegraphics[width=0.9\textwidth]{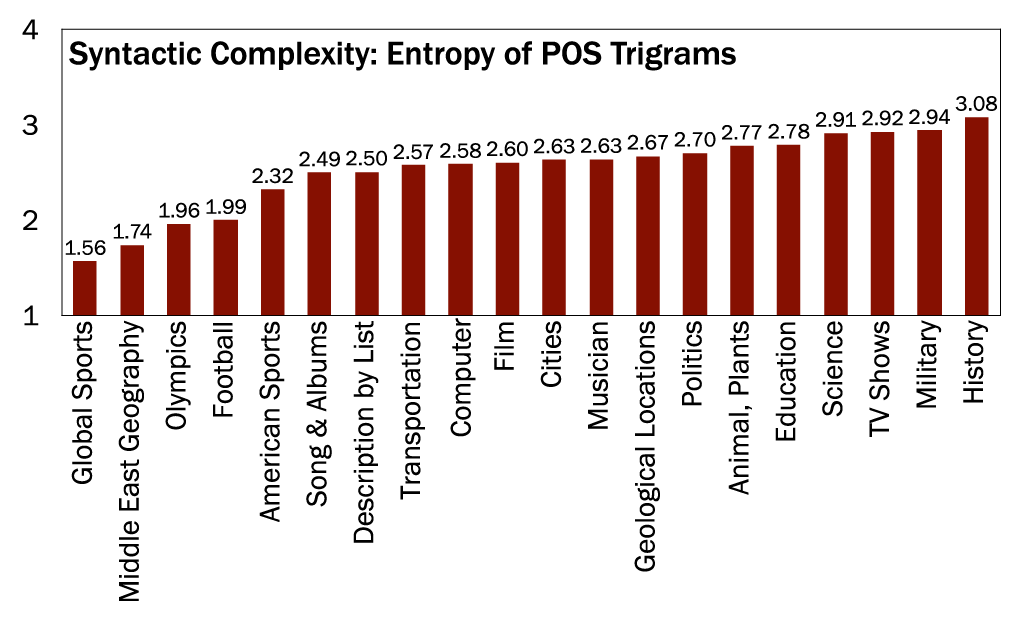}
\vspace{3mm}
\caption{{\bf Language complexity varies by topic.} Language complexity as measured by the entropy of POS trigrams varies by topic. Thus, we control for topic in order to measure language complexity more accurately.}
\label{lc_by_topics}
\vspace{2mm}
\end{figure}

Prior to computing the language complexity measures for edits, we control for the topics of the articles presented in the previous section, as prior research found that the language used in conceptual articles tend to be more complex than the language used in biographical and factual articles~\cite{yasseri2012practical}. Fig. ~\ref{lc_by_topics} shows that the complexity of language differs greatly by topic. Sports-related and other fact-oriented articles (e.g. \texttt{Football}, \texttt{Middle East Geography}, and \texttt{Olympics}) show lower language complexity than conceptual articles (e.g., \texttt{History}, \texttt{Education}, and \texttt{Science}). To control for topic, we calculate each of language complexity measures within a topic (intra-topic) and then average them over all topics (inter-topic). 

\begin{figure}[!htb]
\centering
\includegraphics[width=0.97\textwidth]{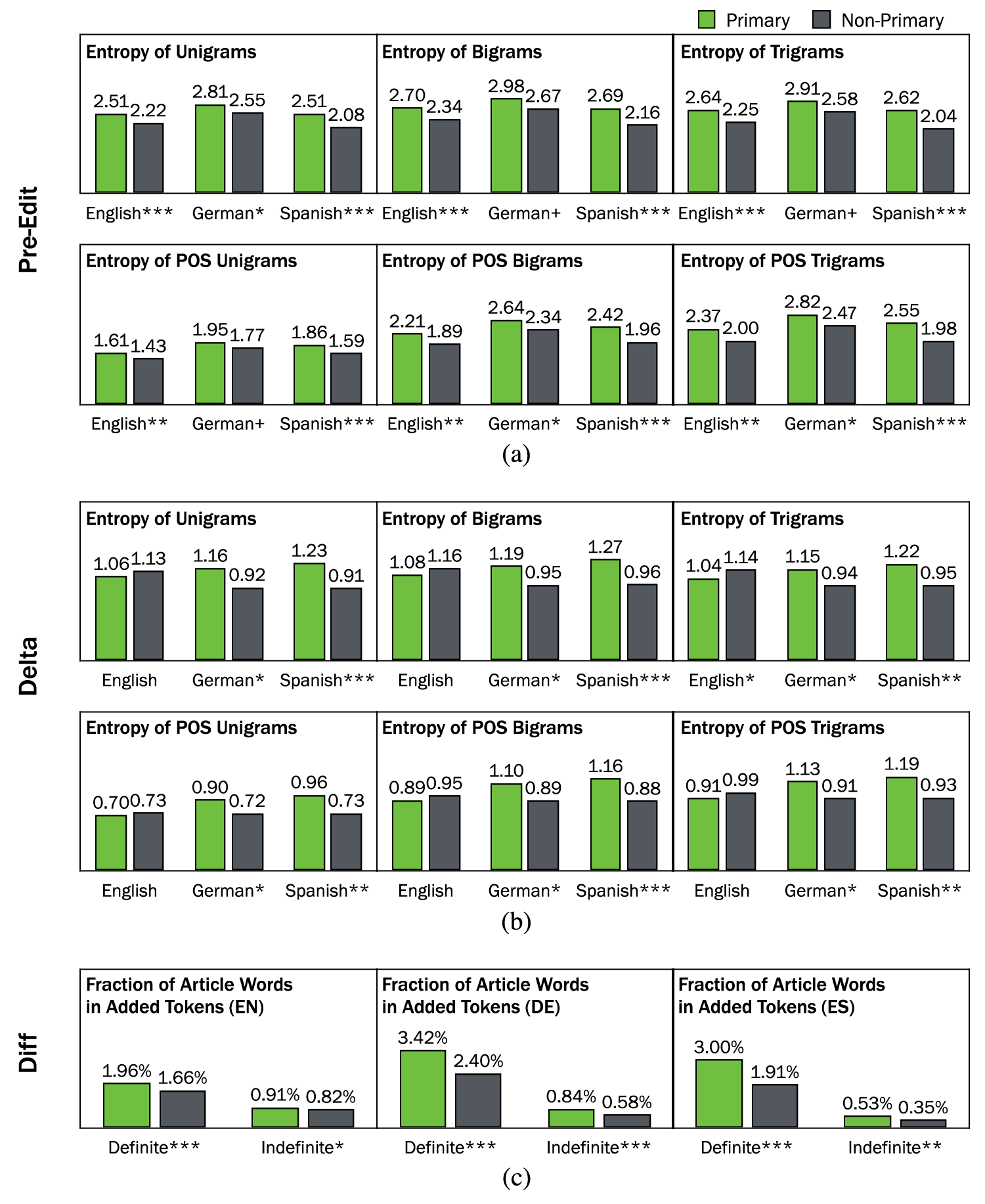}
\caption{(a) {\bf Language complexity of pre-edits for primary and non-primary editors.} Both lexical diversity measures and syntactic diversity measures show primary users edit more complex articles.  (b) {\bf The increase of lexical and syntactic diversity per paragraph per edit (delta).} The higher delta  complexity scores for primary users indicate that multilinguals have higher linguistic abilities and make more complex edits in their primary languages. (c) {\bf Fraction of article words in added tokens.} Primary editors use more article words, both definite and indefinite articles, than non-primary editors. (${}^+p < 0.08$; *$p < 0.05$; **$p < 0.01$; ***$p < 0.001$)}
\label{results_everything}
\end{figure}

\paragraph{Pre-Edit} 
We first examine the complexity of \texttt{Pre-Edits} to understand the text which multilingual editors choose to edit in their primary and non-primary languages. 

Fig.~\ref{results_everything}a shows that the entropy of unigrams, bigrams, and trigrams is always higher for the edits of multilingual users writing in English or Spanish as a primary language compared to those writing in English or Spanish as a non-primary language.
The same is true for unigram entropy in German, while bigram and trigram entropy show a similar pattern although the difference is not significant. In short, we observe a difference between primary and non-primary editors for all conditions, indicating that primary editors use more diverse terms while editing. 

Moreover, we observe that the entropy of part-of-speech unigrams, bigrams, and trigrams is significantly higher for edits by primary editors in English and Spanish. Likewise, the POS entropy of edits is always higher for primary than non-primary editors in the German edition. The difference for bigrams and trigrams is significant while the difference in unigram POS entropy is not. This indicates that syntactic complexity also differs between primary and non-primary editors in all three languages in general. Thus, we conclude that primary editors edit more complex parts of articles compared to non-primary editors. In addition, the result implies that multilingual editors writing in a non-primary language may face a complexity barrier whereby they shy away from editing more complex sections of articles.

\paragraph{Delta} 
Delta complexity measures the difference in complexity before and after each article edit session and thus provides a measure of how the edits by the multilingual user changed the complexity of the article.

Fig.~\ref{results_everything}b shows that there are significant differences between primary and non-primary groups, but that these differences are not consistent across the three language editions. Specifically, for the German and Spanish editions, we find the average increase in entropy between article edit sessions is higher from primary than non-primary editors. Surprisingly, however, we do not observe a significant difference between editors writing in English as a primary or non-primary language for the lexical diversity and syntactic complexity. Opposite to the other language editions, all entropy measures are slightly higher for non-primary editors of the English edition compared to primary editors.

These results implies that primary editors in German and Spanish editions possess higher linguistic proficiency, in terms of writing ability at least, than non-primary editors. This is consistent to the findings from the analysis of user interest. However, this does not apply to English, where there is no evidence that primary and non-primary editors of English possess different levels of linguistic proficiency.

\paragraph{Diff}
Fig.~\ref{results_everything}c shows that there are significant differences in the use of articles between by primary and non-primary editors. For all three language editions, primary editors use more articles---both definite and indefinite---than non-primary editors. 
Given that choosing the proper article is a difficult task---even proficient learners only have an accuracy of $\sim90\%$~\cite{mcenery2006corpus}---the difference in the use of articles between primary and non-primary users likely stems from different levels of proficiency. It is possible that non-primary editors simply skip using article words when confused as to which article to use.

\section{Discussion}
In this article we looked at the editing behaviors of multilinguals in Wikipedia, one of the largest online user-generated content platforms. The results of analyzing engagement shows that multilinguals add more content and spend significantly more time per article edit session in their primary languages. 
In all three language editions, we find that multilingual users are slightly more likely to make non-visible edits that do not change any of the visible article text in their non-primary languages compared to their primary languages. Past work has shown many multilingual users engage with images and multimedia content in their non-primary languages~\cite{hale2015,hale2012}, and a portion of these non-visible edits may be image-related although we have not specifically tested that possibility.

We also find that the overall distribution of interests is very similar between primary and non-primary users in each language although there are a few differences by topic and language. For the few topics with significantly different levels of interest between primary and non-primary users, we find that the topics with more interest from primary users are more complex than the topics with more interest from non-primary users for the German and Spanish editions. We do not observe the same pattern in English, however.

Within the Spanish and German editions, we find that primary users choose to edit more complex text than non-primary users (pre-edit), that the edits of primary users result in a larger increase in the complexity of the articles than the edits of non-primary users (delta), and that the content of the edits primary users show greater language proficiency compared to non-primary users (diff, particularly the use of articles). 

The English edition of Wikipedia shows similar findings for the types of articles primary and non-primary users choose to edit (pre-edit) and for the use of articles (diff). However, the results for the English edition are markedly different from the Spanish and German editions in regard to the effect of edits by multilingual users editing the English edition (delta). We find almost no difference in the effects of edits by multilingual users to English articles regardless as to whether the multilingual is a primary editor of the English edition or a non-primary editor of the edition. If anything, the edits of non-primary users raise the complexity of articles in the English edition slightly more than the edits of primary users.

These findings reinforce how strong of a barrier language is on user-generated content platforms. Even multilingual users who edit multiple editions of Wikipedia devote most of their efforts to editing the edition of their primary language---making more edits and spending more time within article edit sessions in their primary languages than in their non-primary languages.

When multilingual users do edit in a second, non-primary language they often face a language complexity barrier. Users editing a non-primary language edition restricted their edits to less complex parts of articles. In addition, with the exception of English, the edits that they made did not raise the linguistic complexity of the articles as much as the edits by multilingual users who primarily edited the language. This accords with linguistics research that multilinguals have differing levels of competency in their languages and that such competency is often related to how much they use each language~\cite{haugen1969,grosjean2010}.

However, the contribution of editors in English edition is a unique and noteworthy exception to the general pattern. 
In this way, we add to the findings about the unique role English has online as a bridge between multiple languages~\cite{hale2014twitter,hale2014wiki,herring2007,kim2014sociolinguistic}.
Since non-primary editors of the English edition of Wikipedia are able to make equally complex edits as primary editors of the edition, they can better overcome the language barrier and contribute to the English version of documents using knowledge from their primary languages. In this way they might help mitigate the information asymmetries between English and the other language editions, and information may further flow from the English edition into other language editions~\cite{warncke-wang2012}. The potential of English as a bridge language is somewhat muted on Wikipedia, however, given that many foreign-language articles do not an equivalent article in English~\cite{hecht2010babel}. This is partially driven by the interests of editors, but also by differing standards as to what is noteworthy or significant enough to have an article in different languages~\cite{callahan2011,hautasaari2012}.

In this article we have analyzed the complexity of Wikipedia edits by multilingual users in their primary and non-primary languages. One concern is how well a multilingual user's primary language (i.e., the languages of the edition of Wikipedia that the user edits most frequently) maps to their native language. Limited ground truth data is available as only a small fraction of Wikipedia users disclose their native languages on their personal user pages. 
Nevertheless, we checked the alignment between users' primary languages and their self-declared native languages for English by constructing another dataset of all 221,162 editors who contributed to one or more articles in the English edition of Wikipedia appearing within the category ``Wikipedia Controversial Topics.'' Among these editors, there are 18,962 users who disclosed their native languages on their user pages. Since the editors are crawled from the edit histories of the English edition of Wikipedia, we retain only the editors whose primary language or disclosed native language is English. We find that 94.4\% of these editors (1,604 of 1,699) have the same primary language as the disclosed native language, indicating that there is very close alignment between the primary languages and the native languages of the multilingual users.

There remains a range of questions to be tackled. In addition to Wikipedia, there are many other sources for user-generated contents. In addition to the specific editing behaviors we studied here, there are many other behavioral patterns that could be examined to reveal the nature of knowledge diffusion and contributions of multilingual users. Furthermore, to understand the behavior and the roles of multilingual users more fully the contributions of multilingual users should also be compared with those of monolingual users.

\bibliography{library}
\bibliographystyle{plain}

\pagebreak

\section*{Supporting Information}

\subsection*{S1 Topic Clusters from the English Edition of Wikipedia}
\label{appendix:topic_clusters}

\begin{table}[htb!]
\begin{tabular}{l|l}
\hline
\textbf{Topic 1: Science} & \textbf{Topic 2: Football} \\ \hline
Museum Boerhaave & Bebé \\
History of photography & Georgi Kinkladze \\
Planetarium & Raïs M'Bolhi \\
List of Dutch inventions and discoveries & John Carew \\
Electrotyping & John Guidetti \\ \hline

\textbf{Topic 3: Film} & \textbf{Topic 4: Middle East Geography} \\ \hline
Levar Burton	&	Counties of Iran	\\
Tribeca Film Festival	&	 Shevir	\\
Romeo and Juliet (films)	&	 Robat (disambiguation)	\\
Viggo Mortensen	&	 Rijan	\\
Walt Disney	&	 Chalmeh	\\ \hline

\textbf{Topic 5: American Sports} & \textbf{Topic 6: Songs \& Albums} \\ \hline
October 2005 in sports	&	Hurry Up, We're Dreaming	\\
NHL trade deadline	&	 	Paradise (Lana Del Rey EP)\\
2005 Nebraska Cornhuskers baseball team	& The Sweet Escape	\\
Bobby Ryan	&	 Ceremonials	\\
History of the New York Jets	& Do It Again (the Beach Boys Song)	 	\\ \hline 

\textbf{Topic 7: Musicians} & \textbf{Topic 8: Cities} \\ \hline
The Decemberists	&	Schüttorf	\\
The Band	&	 Rodgau	\\
Carlos Santana	&	Aachen	\\
Brian May	&	 Olpe	\\
Dirty Three	&	Lisbon	\\ \hline

\textbf{Topic 9: Global Sports} & \textbf{Topic 10: TV Shows} \\ \hline
List of association football competitions	&	The Hunger Games Trilogy	\\
1982 FIFA World Cup	&	The Penguins of Madagascar	\\
Promotion and relegation	&	The Real Adventures of Jonny Quest	\\
Lokomotiv Cove FC	& The Animatrix	\\
1998 FIFA World Cup	&	 M.I. High	\\ \hline

\end{tabular}
\caption{\bf Examples of Wikipedia Article Titles for Topics 1--10 }
\end{table}

\begin{table}[htb!]
\begin{tabular}{l|l}
\hline
\textbf{Topic 11: Politics} & \textbf{Topic 12: History} \\ \hline
European Conservatives and Reformists	&	Pierre-Marie-Alphonse Favier	\\
Viktor Yushchenko	&	 Society of the Song Dynasty	\\
List of state leaders in 1993	&	 Taukei Ni Waluvu	\\
Tarja Halonen	&	 17th century	\\
United Kingdom Alternative Vote Referendum	&	 Eruera Maihi Patuone	\\ \hline

\textbf{Topic 13: Military} & \textbf{Topic 14: Transportation} \\ \hline
Air Warfare of World War II	&	Railway station layout	\\
Battle of Jutland	& Railways in Melbourne	\\
Defence of the Reich	&	 LSWR suburban lines	\\
Battle of the River Plate	&	 Train station	\\
A World at War	&	 Hastings Line	\\ \hline

\textbf{Topic 15: Computer} & \textbf{Topic 16: Education} \\ \hline
Packet switching	&	Stanford University	\\
Videoconferencing	&	 Waded Cruzado	\\
List of software forks	&	Massachusetts Institute of Technology	\\
Bulletin board system	&	Vishen Lakhiani	\\
Trusted computing	&	University of California, Berkeley	\\ \hline

\textbf{Topic 17: Geographical Locations} & \textbf{Topic 18: Descriptive} \\ \hline
Quehanna Wild Area	&	BBC Sport	\\
Global storm activity of 2009	&	BMC Racing Team	\\
Death Valley National Park	&	Coca-Cola formula	\\
Kimberley (Western Australia)	&	Coal	\\
Hwange National Park	&	Colombian cuisine	\\ \hline

\textbf{Topic 19: Olympics} & \textbf{Topic 20: Animals \& Plants} \\ \hline
Chronological summary of the 2012 Summer Olympics	&	Lobatus Gigas	\\
2013 IPC Athletics World Championships	&	 List of recently extinct birds	\\
2014 in Sports	&	 Galápagos Tortoise	\\
Yelena Isinbayeva	&	 Common Starling	\\
1980 Summer Olympics	&	 Giant Trevally	\\ \hline
\end{tabular}
\caption{ \bf{Examples of Wikipedia Article Titles for Topics 11--20}  }
\end{table}

\clearpage
\subsection*{S2 Topic Clusters from the Spanish Edition of Wikipedia}
\label{appendix:topic_clusters_spanish}

\begin{table}[htb!]
\begin{tabular}{l|l}
\hline
{\bf Topic 1: Art}                         & {\bf Topic 2: Descriptive}                    \\ \hline
Kinetic art                                & Ancient history                               \\
Self-portrait                              & Psychokinesis                                 \\
Claude Monet                               & European Train Control System                 \\
History of architecture                    & Functional programming                        \\
São Paulo Museum of Art                    & Pluto                                         \\\hline
{\bf Topic 3: Soccer}                      & {\bf Topic 4: Film}                           \\\hline
Celtic F.C.                                & The Great Gatsby (1974 film)                  \\
Carlos Tévez                               & Tom Hiddleston                                \\
1964 Argentine Primera División            & Nickelodeon Movies                            \\
Everton F.C.                               & Penélope Cruz                                 \\
FC Barcelona                               & Johnny Depp                                   \\
Kaley Cuoco                                & Kaley Cuoco                                   \\\hline
{\bf Topic 5: Animal}                      & {\bf Topic 6: Global Sports}                  \\\hline
Nudochernes leleupi                        & Asian Football Confederation                  \\
Oxydactylus                                & 2013 FIFA Confederations Cup                  \\
Mordellistena chopardi                     & 2018 FIFA World Cup                           \\
Parazaona pycta                            & UEFA Women's Euro 2013                        \\
Desert Sparrow                             & Argentina national rugby union team           \\\hline
{\bf Topic 7: History}                     & {\bf Topic 8: Plants}                         \\\hline
Gustaf VI Adolf of Sweden                  & Notholaena                                    \\
Hundred Years' War                         & Martretia                                     \\
King in Prussia                            & Peach                                         \\
Catholic Monarchs                          & Trevo                                         \\
Maximilian I of Mexico                     & Totora (plant)                                \\\hline
{\bf Topic 9: Politicians}                 & {\bf Topic 10: Natural Science}               \\\hline
Alfonso Guerra                             & Ozone layer                                   \\
List of Governors of Iowa                  & Nitrogen cycle                                \\
First Lady of Peru                         & Biomechanics                                  \\
Ban Ki-moon                                & Earth's rotation                              \\
President of Honduras                      & Spinal cord                                   \\\hline
\end{tabular}
\caption{ \bf{Examples of Spanish Wikipedia Article Titles for Topics (presented in English) 1--10}  }
\label{Spanish_topic_ex_1}
\end{table}

\begin{table}[htb!]
\begin{tabular}{l|l}
\hline
{\bf Topic 11: Social Science}             & {\bf Topic 12: Music}                         \\\hline
Democratic centralism                      & Red (Taylor Swift album)                      \\
Political economy                          & Boys Like Girls                               \\
Market research                            & Barcelona (song)                              \\
Social movement                            & Get Lucky (Daft Punk song)                    \\
The Third Wave                             & Folk music                                    \\\hline
{\bf Topic 13: Cities}                     & {\bf Topic 14: Geographical Locations}          \\\hline
Machu Picchu                               & Regions of Senegal                            \\
Madrid                                     & Guavio Province                               \\
New Delhi                                  & Central Region Venezuela                      \\
Montreal                                   & Province of Castellón                         \\
Santiago                                   & Khar Turan National Park                      \\\hline
{\bf Topic 15: Olympics}                   & {\bf Topic 16: Literature}                    \\\hline
Czechoslovakia at the 1988 Winter Olympics & Spanish Golden Age                            \\
Denmark at the 1992 Winter Olympics        & Voltaire                                      \\
Great Britain at the 1992 Winter Olympics  & Poetry                                        \\
Finland at the 2010 Winter Olympics        & Franz Kafka                                   \\
Hungary at the 1956 Winter Olympics        & Giovanni Papini                               \\\hline
{\bf Topic 17: Musicians}                  & {\bf Topic 18: Politics}                      \\\hline
Metallica                                  & Venezuelan Democratic Party                   \\
Billy Joel                                 & Social Christian Reformist Party              \\
Janis Joplin                               & Andalusian parliamentary election 2004        \\
Keith Emerson                              & United States presidential election 1788–1789 \\
Marilyn Manson                             & Green Ecologist Party (Chile)                 \\\hline
{\bf Topic 19: Entertainment}              & {\bf Topic 20: Tennis}                        \\\hline
Half-Life 2                                & 2013 ATP Challenger Tour                      \\
Glee (TV series)                           & Canadian Open (tennis)                        \\
The Phantom of the Opera (1986 musical)    & 2013 Western \& Southern Open                 \\
Garfield                                   & Maria Sharapova                               \\
Nickelodeon                                & Rafael Nadal                                  \\\hline
\end{tabular}
\caption{ \bf{Examples of Spanish Wikipedia Article Titles for Topics (presented in English) 11--20}  }
\label{Spanish_topic_ex_2}
\end{table}

\clearpage
\subsection*{S3 Topic Clusters from the German Edition of Wikipedia}

\begin{table}[htb!]
\begin{tabular}{l|l}
\hline
{\bf Topic 1: Computer}           & {\bf Topic 2: Natural Science}              \\\hline
IPv6 rapid deployment             & Boiling point                               \\
C (programming language)          & Solar cell                                  \\
Real-time operating system        & Recombinant DNA                             \\
Graphical user interface          & Klinefelter syndrome                        \\
Web Server Gateway Interface      & Volcano                                     \\\hline
{\bf Topic 3: Descriptive}        & {\bf Topic 4: Geographical Loc.~(Locations): U.S.}  \\\hline
Abraham Lincoln                   & Moorhead Mississippi                        \\
Anti-Zionism                      & Madison County Texas                        \\
Ice                               & List of cities in Virginia                  \\
Eiffel Tower                      & Leon County Florida                         \\
BMW 340                           & San Rafael California                       \\\hline
{\bf Topic 5: Names}              & {\bf Topic 6: Geographical Loc.~(Locations): Europe} \\\hline
Watson (Familienname)             & Westfalen                                   \\
Williamson (Surname)              & Pram Austria                                \\
Taylor (Name)                     & Monforte Portugal                           \\
Shaw (Name)                       & Hamburg-Marienthal                          \\
Peter (Given Name)                & Cologne (Region)                            \\\hline
{\bf Topic 7: History}            & {\bf Topic 8: Academic}               \\\hline
List of state leaders in 655      & University of Mannheim                      \\
Charles VIII of Sweden            & Computer science                            \\
Frederick William I of Prussia    & Individual psychology                       \\
Vianden Castle                    & Industrial sociology                        \\
Guelderian Wars                   & Keynesian economics                         \\\hline
{\bf Topic 9: Celebrities}        & {\bf Topic 10: Soccer}                      \\\hline
List of American novelists        & List of football clubs in Germany           \\
Lists of golfers                  & 2003–04 UEFA Cup                            \\
List of Medal of Honor recipients & Real Madrid C.F.                            \\
Franklin Roosevelt                & Lionel Messi                                \\
Sebastian Shaw                    & Son Heung-Min                               \\\hline                    \end{tabular}
\caption{ \bf{Examples of German Wikipedia Article Titles for Topics (presented in English) 1--10}  }
\label{German_topic_ex_1}
\end{table}

\begin{table}[htb!]
\begin{tabular}{l|l}
\hline
{\bf Topic 11: Cultural Heritage}                   & {\bf Topic 12: Musicians}                    \\\hline
Liste der Kulturgüter in Oberwil-Lieli              & Eminem                                       \\
Liste der denkmalgeschützten Objekte in Krumbach    & Flo Rida                                     \\
Liste der denkmalgeschützten Objekte in Gleisdorf   & 2012 Grammy Awards                           \\
Liste der denkmalgeschützten Objekte in Arnoldstein & John Lennon                                  \\
Liste der denkmalgeschützten Objekte in Bürs        & Lady Gaga                                    \\\hline
{\bf Topic 13: Natural Topography}                  & {\bf Topic 14: Land Transport}               \\\hline
Najerilla (River)                                   & Hesper Valley Railway                        \\
Olympic National Park                               & Innsbruck Central Station                    \\
Nock Mountains                                      & Highways in Bulgaria                         \\
Niedersonthofener See                               & List of tunnels in Germany                   \\
Quincy Bay                                          & Metrobus                                     \\\hline
{\bf Topic 15: Politicians}                         & {\bf Topic 16: Entertainment}                \\\hline
Rudolf Hierl (Politiker)                            & Denzel Washington                            \\
Ulrich Kelber                                       & Bradley Cooper                               \\
Robert Thaller                                      & Anna Karenina (2012 Film)                    \\
Norbert Otto (Politiker)                            & Braid (Video Game)                           \\
Liam Aylward                                        & Boss (TV series)                             \\\hline
{\bf Topic 17: Air transport}                       & {\bf Topic 18: Global Sports}                \\\hline
Lombok International Airport                        & Badminton at the Asian Games                 \\
Salzburg Airport                                    & Australian Football International Cup        \\
Airline                                             & Australian Goldfields Open                   \\
Aviation accidents and incidents                    & Australia national association football team \\
Human-powered helicopter                            & 2012–13 FA Cup                               \\\hline
{\bf Topic 19: Authors}                             & {\bf Topic 20: Military}                     \\\hline
Franz Kafka                                         & Japanese conquest of Burma                   \\
Jan Zimmermann                                      & Pacific War                                  \\
Herbert Schmidt                                     & Russian Navy                                 \\
List of authors by name: F                          & Soviet submarine K-19                        \\
List of authors by name: W                          & Rocket artillery                             \\\hline
\end{tabular}
\caption{ \bf{Examples of German Wikipedia Article Titles for Topics (presented in English) 11--20}  }
\label{German_topic_ex_2}
\end{table}

\clearpage
\subsection*{S4 Syntactic Complexity for each topics by Three Language Editions}

\begin{table}[!htb]
\centering
    \begin{tabular}{l|r|r|r|r|r}
    \hline
Topic & Complexity & Primary & Non-Primary & t-value & p-value \\
\hline
\hline
Science & 2.917 & 0.064 & 0.061 & 0.747 & 4.550E-01 \\
Football & 1.995 & 0.029 & 0.031 & -0.756 & 4.496E-01 \\
Film & 2.610 & 0.046 & 0.048 & -0.542 & 5.880E-01 \\
Middle East Geography & 1.733 & 0.001 & 0.002 & -1.081 & 2.799E-01 \\
American Sports & 2.312 & 0.006 & 0.007 & -0.361 & 7.184E-01 \\
Music: Song, Albums & 2.495 & 0.030 & 0.026 & 1.343 & 1.793E-01 \\
Music: Musician & 2.630 & 0.057 & 0.060 & -0.801 & 4.231E-01 \\
Cities & 2.638 & 0.043 & 0.057 & -4.023 & 5.796E-05 \\
Sports Related Articles & 1.563 & 0.047 & 0.048 & -0.122 & 9.029E-01 \\
TV Shows & 2.930 & 0.056 & 0.059 & -0.840 & 4.009E-01 \\
Politics & 2.701 & 0.030 & 0.027 & 0.955 & 3.396E-01 \\
History & 3.071 & 0.196 & 0.189 & 1.208 & 2.271E-01 \\
Military & 2.938 & 0.033 & 0.034 & -0.536 & 5.919E-01 \\
Transportation & 2.565 & 0.014 & 0.016 & -1.368 & 1.715E-01 \\
Computer & 2.577 & 0.057 & 0.049 & 2.099 & 3.582E-02 \\
Education & 2.776 & 0.078 & 0.074 & 0.846 & 3.975E-01 \\
Geographical Locations & 2.676 & 0.023 & 0.019 & 1.753 & 7.960E-02 \\
Descriptive & 2.505 & 0.152 & 0.151 & 0.129 & 8.974E-01 \\
Olympics & 1.952 & 0.022 & 0.023 & -0.348 & 7.278E-01 \\
Animal, Plants & 2.770 & 0.017 & 0.020 & -1.471 & 1.413E-01 \\
    \hline
    \end{tabular}
\caption{{\bf Syntactic Complexity for 20 English topics}.}
\end{table}

\begin{table}[!htb]
\centering
    \begin{tabular}{l|r|r|r|r|r}
    \hline
Topic & Complexity & Primary & Non-Primary & t-value & p-value \\
\hline
\hline
Computer & 3.219 & 0.047 & 0.030 & 3.070 & 2.157E-03 \\
Natural Science & 3.511 & 0.035 & 0.022 & 2.943 & 3.276E-03 \\
Descriptive & 3.220 & 0.304 & 0.322 & -1.445 & 1.486E-01 \\
Geographical Loc.: U.S. & 2.025 & 0.002 & 0.002 & -0.434 & 6.643E-01 \\
Names & 2.361 & 0.013 & 0.011 & 0.700 & 4.840E-01 \\
Geographical Loc.: Europe & 2.964 & 0.036 & 0.028 & 1.685 & 9.200E-02 \\
History & 3.341 & 0.046 & 0.037 & 1.619 & 1.055E-01 \\
Academic & 3.432 & 0.123 & 0.117 & 0.709 & 4.783E-01 \\
Celebrities & 3.120 & 0.038 & 0.036 & 0.392 & 6.949E-01 \\
Soccer & 2.769 & 0.036 & 0.051 & -2.393 & 1.679E-02 \\
Cultural Heritage & 1.906 & 0.004 & 0.001 & 2.404 & 1.629E-02 \\
Musicians & 3.034 & 0.049 & 0.053 & -0.483 & 6.292E-01 \\
Natural Topography & 3.062 & 0.025 & 0.025 & 0.019 & 9.845E-01 \\
Land Transport & 3.306 & 0.026 & 0.027 & -0.139 & 8.896E-01 \\
Politicians & 3.207 & 0.056 & 0.067 & -1.756 & 7.921E-02 \\
Entertainment & 3.331 & 0.052 & 0.061 & -1.408 & 1.591E-01 \\
Air Transport & 3.291 & 0.017 & 0.016 & 0.194 & 8.461E-01 \\
Global Sports & 2.531 & 0.024 & 0.035 & -2.173 & 2.987E-02 \\
Authors & 3.429 & 0.046 & 0.037 & 1.765 & 7.760E-02 \\
Military & 3.630 & 0.021 & 0.021 & -0.050 & 9.603E-01 \\
    \hline
    \end{tabular}
\caption{{\bf Syntactic Complexity for 20 German topics}.}

\end{table}

\begin{table}[!htb]
\centering
    \begin{tabular}{l|r|r|r|r|r}
    \hline
Topic & Complexity & Primary & Non-Primary & t-value & p-value \\
\hline
\hline
Art & 3.012 & 0.016 & 0.016 & 0.025 & 9.804E-01 \\
Descriptive & 2.808 & 0.384 & 0.418 & -2.159 & 3.092E-02 \\
Soccer & 2.347 & 0.077 & 0.072 & 0.593 & 5.536E-01 \\
Film & 2.747 & 0.038 & 0.038 & 0.064 & 9.494E-01 \\
Animal & 2.444 & 0.002 & 0.002 & -0.358 & 7.201E-01 \\
Global Sports & 2.295 & 0.038 & 0.031 & 1.115 & 2.648E-01 \\
History & 3.001 & 0.011 & 0.010 & 0.417 & 6.769E-01 \\
Plants & 2.376 & 0.005 & 0.017 & -3.429 & 6.189E-04 \\
Politicians & 3.179 & 0.032 & 0.019 & 2.443 & 1.466E-02 \\
Natural Science & 3.084 & 0.039 & 0.051 & -1.676 & 9.387E-02 \\
Social Science & 3.005 & 0.042 & 0.030 & 2.023 & 4.322E-02 \\
Music & 2.789 & 0.052 & 0.039 & 1.741 & 8.183E-02 \\
Cities & 2.931 & 0.081 & 0.087 & -0.670 & 5.029E-01 \\
Geographical Locations & 2.523 & 0.020 & 0.037 & -3.324 & 8.991E-04 \\
Olympics & 1.970 & 0.001 & 0.002 & -0.423 & 6.725E-01 \\
Literature & 3.094 & 0.033 & 0.037 & -0.645 & 5.193E-01 \\
Musicians & 2.814 & 0.027 & 0.024 & 0.503 & 6.147E-01 \\
Politics & 3.029 & 0.022 & 0.019 & 0.599 & 5.492E-01 \\
Entertainment & 2.967 & 0.076 & 0.046 & 3.736 & 1.921E-04 \\
Tennis & 2.654 & 0.004 & 0.005 & -0.552 & 5.811E-01 \\
    \hline
    \end{tabular}
\caption{{\bf Syntactic Complexity for 20 Spanish topics}.}
\end{table}

\end{document}